\documentclass[amsmath,amssymb,showpacs,lengthcheck,reprint,floatfix]{achemso} 
\usepackage[T1]{fontenc}
\usepackage[latin9]{inputenc}
\usepackage{mathrsfs}
\usepackage{graphicx,amsmath,amssymb}
\usepackage[usenames,dvipsnames]{color}

			    \definecolor{mygray}{gray}{0.88}
     \newcommand{\note}[1]{{#1}}

\newcommand{\epstd}{\tilde\varepsilon}
    
\usepackage{braket}
\usepackage{placeins}

\usepackage{blindtext}

\title{An anisotropic functional for two-dimensional material systems}  
\author{Michael Lorke}
\email{michael.lorke@uni-due.de}
\affiliation{Faculty of Physics, University of Duisburg-Essen, Germany}

\begin{document}
\begin{abstract}
\note{
Density functional theory is the workhorse of modern electronic structure theory. However, its accuracy in practical calculations is limited by the choice of the exchange-correlation functional. In this respect, two-dimensional materials pose a particular challenge because their electronic structure is characterized by strong spatial inhomogeneity. This spatial inhomogeneity is not adequately captured by conventional exchange-correlation functionals, which are predominantly based on isotropic descriptions of electronic interactions. We present an anisotropic screened-exchange functional that addresses this limitation and reproduces the band gaps of 2D materials as well as the piecewise linearity of the total energy with fractional occupation.
}
\end{abstract}
\maketitle

\section{Introduction}
\label{sec:introduction}
Almost all applications of semiconductor technology in today's world are based on nanotechnology,
either to create new functionalities or to reduce power consumption as a contribution to global energy savings.
Two-dimensional (2D) materials, defined here as atomically thin crystalline systems with electronic motion confined to a plane and weak out-of-plane dielectric screening, exhibit electronic and optical properties that differ qualitatively from those of their three-dimensional (3D) bulk counterparts. 

The field of 2D materials has created a rich research environment for the development of novel electronic and optical devices.
In this field, the focus is currently shifting from investigations of pure 2D materials themselves to functionalized materials, e.g. by defects in MoS$_2$\cite{klein2019site,mitterreiter2021role,hoetger2021gate},  or hBN\cite{Lopez-Morales:20} for single-photon emitters.
Reduced dimensionality leads to strongly nonlocal and anisotropic dielectric screening, which profoundly affects quasi-particle band gaps, excitonic binding energies, and the localization of defect states. An accurate first-principles description of dielectric screening is therefore a central requirement for predictive simulations of electronic and optical properties in 2D materials.

Within density functional theory (DFT), semilocal exchange--correlation functionals such as the generalized gradient approximation (GGA) are known to underestimate band gaps and to exhibit spurious convexity of the total energy as a function of fractional electron occupation \cite{Perdew:81,Baumeier:06}. These deficiencies are particularly pronounced in 2D systems, where reduced dielectric screening enhances self-interaction errors and artificial charge delocalization. Hybrid functionals that incorporate a fraction of nonlocal Hartree--Fock (HF) exchange partially alleviate these issues \cite{Deak:10,Han:17,Deak:17,Deak:19}; however, widely used hybrids such as PBE0 and Heyd-Scuseria-Ernzerhof (HSE) were designed for bulk materials and employ either unscreened or empirically short-range-screened Coulomb interactions \cite{Becke:96,PBE0,HSE,Krukau:08,Zheng:11,Alkauskas:11,Skone:14,Chen:18}. 
These are often unable to provide an accurate description \cite{Deak19slab}, as
they fail to reproduce the correct asymptotic behavior of dielectric screening.
Numerous GW studies have demonstrated substantial band-gap underestimation
and incorrect defect level alignment when using semilocal or conventional
hybrid functionals in 2D materials such as MoS$_2$ and related compounds, see, e.g., Refs.~\cite{Ramasubramaniam:12,Qiu:13,Komsa:13}.

While GW or related microscopic methods are available for the electronic structure, 
these at present are computationally prohibitive for large supercell models. Moreover, forces and geometry optimizations at the GW level are not routinely available in current first-principles implementations and remain computationally prohibitive for large supercell calculations.
A class of functionals explicitly constructed to satisfy Koopmans-like conditions \cite{Perdew:81SI,Dabo:10,Nguyen:18} has been developed, offering a route to inherently gKT-compliant approximations. As their broader application to solids is still being developed,
we will pursue a different route in this work, namely the development of approximate screened-exchange approaches within the Kohn--Sham DFT framework, aiming to capture the dominant physics of nonlocal exchange at a substantially reduced computational cost.

In this context, a screened-exchange functional for bulk materials was introduced in Ref.~\cite{Lorke20Koopmans}, where the bare Coulomb interaction in the Fock operator is replaced by a statically screened interaction characterized by an isotropic macroscopic dielectric constant. While this approach has proven successful for bulk semiconductors, it does not account for the intrinsically anisotropic and nonlocal nature of dielectric screening in 2D systems. Separately, a macroscopic dielectric screening framework for layered and 2D materials was developed in Ref.~\cite{rosner_wannier_2015}, providing a physically grounded description of wave-vector-dependent in-plane screening in the long-wavelength limit.

In this work, we propose a screened-exchange functional that significantly improves the description of quasi-particle band gaps of representative 2D materials, 
yields near piecewise-linear behavior of the total energy with respect to fractional occupation numbers, indicating that the functional approximately fulfills the generalized Koopmans theorem (gKT) \cite{Janak:78,Perdew:87,Perdew:97a}.
Approximate piecewise linearity has long been recognized as an important criterion for reducing delocalization errors in modern density-functional approximations, including hybrid functionals \cite{MoriSanchez:08,Lany:09,Cohen:12}.
It also provides a consistent starting point for optical calculations using time-dependent DFT (TDDFT) approaches. By retaining the computational efficiency of hybrid-functional calculations while incorporating physically motivated anisotropic screening, the present method offers a practical and predictive framework for first-principles simulations of electronic and optical properties in 2D materials.

The central contribution of the present work is the formulation of a 
self-consistent anisotropic screened-exchange exchange--correlation functional 
for \note{thin} two-dimensional materials. In contrast to conventional hybrid and 
screened-exchange approaches, which rely on isotropic and often 
wave-vector-independent screening, the present method incorporates a 
wave-vector-dependent, anisotropic macroscopic dielectric function directly 
into the nonlocal exchange operator within the Kohn--Sham framework. 

This construction leads to a qualitatively different effective interaction 
that reproduces the correct long-wavelength limit of dielectric screening in 
two-dimensional systems, while remaining computationally comparable to hybrid 
functionals. At the same time, the formulation provides access to total 
energies, forces, and structural optimization, which are not available in 
standard GW-based approaches.

The present approach therefore establishes a physically motivated and 
computationally efficient framework that bridges the gap between 
first-principles many-body methods and density-functional approximations for 
two-dimensional materials.
To our knowledge, this represents the first self-consistent Kohn--Sham
screened-exchange functional incorporating anisotropic, in-plane wave-vector-dependent
dielectric screening specifically tailored for 2D materials.

\section{Anisotropic screened-exchange functional\label{sec:theory}}

\subsection{Screened-exchange formalism}

Hybrid exchange--correlation functionals improve upon semilocal density functional approximations by incorporating nonlocal Hartree--Fock (HF) exchange, thereby reducing self-interaction errors and restoring a more linear dependence of the total energy on fractional occupation numbers. A physically motivated refinement replaces the bare Coulomb interaction in the Fock operator by a statically screened interaction, yielding a screened-exchange (SX) functional that mimics the dominant exchange physics of many-body perturbation theory at reduced computational cost.

In bulk materials, screened-exchange functionals based on an isotropic macroscopic dielectric constant have proven successful~\cite{Lorke20Koopmans}. However, this approximation breaks down in 2D systems, where dielectric screening is intrinsically anisotropic and nonlocal. In particular, the long-range Coulomb interaction is governed by wave-vector-dependent in-plane screening, while out-of-plane screening remains weak. Capturing this behavior requires an explicit incorporation of anisotropic dielectric screening into the nonlocal exchange operator.

The screened-exchange contribution to the exchange--correlation energy is written as
\begin{equation}
E_{\mathrm{SX}}
=
-\frac{1}{2}
\sum_{i,j}^{\mathrm{occ}}
\int d\mathbf{r} \int d\mathbf{r}'
\,
\psi_i^{*}(\mathbf{r}) \psi_j(\mathbf{r})
W(\mathbf{r},\mathbf{r}')
\psi_j^{*}(\mathbf{r}') \psi_i(\mathbf{r}'),
\label{eq:Esx}
\end{equation}
where $\{\psi_i\}$ are the occupied Kohn--Sham orbitals and $W(\mathbf{r},\mathbf{r}')$ is a statically screened Coulomb interaction.

Functional differentiation yields the corresponding nonlocal exchange potential,
\begin{equation}
\hat{V}_{\mathrm{SX}} \psi_i(\mathbf{r})
=
- \sum_{j}^{\mathrm{occ}}
\psi_j(\mathbf{r})
\int d\mathbf{r}'
\,
W(\mathbf{r},\mathbf{r}')
\psi_j^{*}(\mathbf{r}') \psi_i(\mathbf{r}').
\label{eq:Vsx}
\end{equation}
The accuracy of the screened-exchange functional therefore critically dependents on the quality of the screened interaction $W$.
In the following, we use the equivalent notation
$V^{\mathrm{eff}}(\mathbf{q}) \equiv W(\mathbf{q})$
to emphasize its role as an effective screened interaction entering the
exchange potential.
The nonlocal, wave-vector-dependent nature of dielectric screening in
2D systems has been rigorously derived from first-principles
many-body theory, showing that the screened Coulomb interaction follows
$W(q) = 2\pi / [q\,\varepsilon(q)]$ with a nonlocal dielectric function
$\varepsilon(q)$ \cite{Cudazzo:11,Andersen:15}.
This behavior fundamentally differs from bulk systems and provides the
physical foundation for the present anisotropic screened-exchange functional.

\subsection{Macroscopic dielectric screening in 2D materials }
In the isotropic screened-exchange functional of Ref.~\cite{Lorke20Koopmans},
the screened interaction depends only on the magnitude of the wave vector,
\begin{equation}
W_\mathrm{iso}(\mathbf q)
=
\frac{4\pi}{q^2\varepsilon(q)} ,
\end{equation}
where all spatial directions are screened equivalently.
For a two-dimensional material this approximation is no longer valid.

In contrast to the isotropic bulk case, the dielectric response of a
two-dimensional material cannot be represented by a scalar macroscopic
dielectric constant. Screening parallel and perpendicular to the layer are
fundamentally different, and the finite extension of the electronic density
perpendicular to the layer has to be taken into account.
For a strictly two-dimensional electronic system, the relevant interaction 
entering the exchange operator is the interaction between charges confined 
to the layer. The anisotropic dielectric response can therefore be mapped 
onto an effective in-plane screened interaction by solving the electrostatic 
boundary-value problem of a finite-thickness dielectric slab embedded between 
two surrounding media.
Using the effective interaction for layered systems derived in
Refs.~\cite{Emelyanenko:08,rosner_wannier_2015}, this yields


%
\begin{equation}
V^{\mathrm{2D}}_{\mathrm{eff}}(q)
=
\frac{e^2 F(q)}
{2 \tilde{V} q \varepsilon^{\mathrm{2D}}_{\mathrm{eff}}(q)},
\label{eq:V2Deff}
\end{equation}
where $\tilde{V}$ denotes the supercell volume used for the normalization of
the Coulomb interaction and
\begin{equation}
F(q)
=
\frac{2}{\pi} \arctan \frac{\pi}{q h}
\end{equation}
accounts for the finite spatial extent of the electronic density perpendicular
to the layer, described by the effective layer thickness $h$.
Starting from Eq.~\eqref{eq:V2Deff}, $q$ denotes the magnitude of the
in-plane wave vector, $q=|\mathbf{q}_{\parallel}|$, for brevity.

It is important to emphasize that this transformation does not restore
isotropic screening. Instead, the anisotropic dielectric response and the
finite extension of the electronic density perpendicular to the layer are
contained in the effective in-plane wave-vector dependent dielectric function
$\varepsilon^\mathrm{2D}_\mathrm{eff}(q)$ and form factor $F(q)$.
The central step of the present work is the replacement of the isotropically
screened interaction entering the screened-exchange operator by
$V^{\mathrm{2D}}_{\mathrm{eff}}(q)$, thereby incorporating the anisotropic
dielectric response directly into the self-consistent exchange--correlation
functional.
While the effective dielectric interaction itself follows from the
electrostatic description of anisotropic layered systems
\cite{Emelyanenko:08,rosner_wannier_2015}, these approaches provide the
screened Coulomb interaction rather than an exchange--correlation functional.
Here, this interaction is used as the physical screening entering the
nonlocal exchange operator, resulting in a self-consistent
anisotropic screened-exchange functional within Kohn--Sham DFT.

The effective interaction entering the screened-exchange functional is determined by the 
two-dimensional dielectric screening function
\begin{equation}
\varepsilon^{\mathrm{2D}}_{\mathrm{eff}}(q) = 
\frac{\varepsilon_2(q)\left[1 - \tilde{\varepsilon}_1(q)\tilde{\varepsilon}_3(q)e^{-2qh}\right]}
{1 + \left[\tilde{\varepsilon}_1(q)+\tilde{\varepsilon}_3(q)\right]e^{-qh}
+ \tilde{\varepsilon}_1(q)\tilde{\varepsilon}_3(q)e^{-2qh}},
\label{eqn:epsmodel}
\end{equation}
with
\[
\tilde{\varepsilon}_i(q)=
\frac{\varepsilon_2(q)-\varepsilon_i}{\varepsilon_2(q)+\varepsilon_i},
\qquad i=1,3.
\]
Here, $\varepsilon_1$ and $\varepsilon_3$ denote the dielectric constants of the substrate and cover layer, respectively, while $\varepsilon_2(q)$ is the wave-vector-dependent dielectric function of the monolayer material.

Equation~\eqref{eqn:epsmodel} provides the anisotropic, nonlocal screening
entering the effective interaction and therefore constitutes the key physical
input used to construct the present screened-exchange functional.
For the dielectric response of the monolayer material, $\varepsilon_2(q)$, we employ the model dielectric function previously introduced for bulk materials~\cite{Lorke20Koopmans},
\begin{equation}
\varepsilon_2^{-1}(q)
=
1+\left(\frac{1}{\varepsilon_{b}}-1\right)\frac{1}{\cosh(q/\sigma)},
\label{eq:eps1}
\end{equation}
which satisfies the correct asymptotic limits at small and large wave vectors~\cite{Banyai:98,Gartner:00,Gartner:02}. 

The screening length $\sigma$ follows from a static random-phase approximation
(RPA) to the dielectric function \cite{Cappellini:93,Shimazaki:08}. This yields
\begin{equation}\label{eq:screening_length}
 \sigma=\frac{Z}{\log(2+\sqrt{3})}
 \sqrt{k_{\mathrm{TF}}^2
 \left(
 \frac{1}{\varepsilon_b-1}+1
 \right)},
\end{equation}
where $Z$ denotes a renormalization factor. The Thomas--Fermi wave vector is
given by \cite{Cappellini:93,Shimazaki:08,Shimazaki:10}
\begin{equation}\label{eq:screening_k}
k_{\mathrm{TF}}
=
4
\left(
\frac{3N_{\mathrm{el}}}
{\pi\tilde V}
\right)^{1/3},
\end{equation}
with the volume $\tilde V$.

While the renormalization factor $Z$ is employed as in Ref.~\cite{Lorke20Koopmans}, its value is not reoptimized for the two-dimensional system. Instead, $Z$ is determined from the corresponding bulk material using the three-dimensional screened-exchange functional and subsequently kept fixed for the monolayer calculation. As illustrated in Fig.~\ref{fig1}, 
the intrinsic screening of the monolayer contributes primarily in an intermediate momentum range, whereas the long-wavelength screening is governed predominantly by the dielectric environment. Consequently, the influence of the renormalization factor on the screened interaction is reduced compared to the corresponding three-dimensional functional, where the bulk screening determines the interaction over the entire momentum range. A more detailed discussion of the renormalization factor $Z$ and its properties regarding multiple 3D semiconductor systems is given in Ref.~\cite{Lorke20Koopmans}.
The parameter $\varepsilon_b$ is likewise not an adjustable fitting parameter but is obtained from a PBE calculation of the dielectric response of the corresponding bulk material. As in the three-dimensional screened-exchange functional, this dielectric screening is subsequently kept fixed throughout the calculation rather than being updated self-consistently, following a philosophy similar to that underlying the GW$_0$ approximation.

In the limit of homogeneous isotropic three-dimensional screening, the
present formulation recovers the bulk screened-exchange functional of
Ref.~\cite{Lorke20Koopmans}. The approach therefore represents a direct
extension of screened exchange from isotropic bulk systems to materials with
anisotropic dielectric screening in reduced dimensionality.

\subsection{Relation to GW and scope of the functional}

The present functional belongs to the class of screened-exchange density functionals
\cite{Clark:10,Clark:10a,Lorke20Koopmans}, which are motivated by the static screened-exchange limit of many-body perturbation theory. It can therefore be viewed as a static approximation to the screened-exchange (SEX) contribution of the GW self-energy. Unlike a density-functional implementation of the static COHSEX approximation, the present approach replaces only the exchange part by a screened nonlocal interaction, while retaining semilocal correlation and neglecting the frequency dependence of the self-energy.

\note{The present work extends the screened-exchange framework of Ref.~\cite{Lorke20Koopmans} by replacing the isotropic bulk screening with an anisotropic effective dielectric screening appropriate for thin two-dimensional materials, with dielectric screening from the surrounding environment incorporated through the effective interaction.
The dielectric environment therefore enters through the effective screened interaction, allowing substrate and cover-layer screening to be represented within the screened-exchange framework.
The functional is intended for thin two-dimensional materials whose dielectric screening can be described by the effective quasi-two-dimensional interaction employed here. It is therefore applicable to free-standing as well as environmentally screened systems. Explicitly treated multilayer systems may introduce additional nonlocal interlayer correlation effects that are not contained in the present semilocal correlation treatment. A systematic assessment of the importance of these effects for the applicability of the present functional to multilayer systems is left for future work.
}

By embedding the physically motivated anisotropic dielectric screening directly into the nonlocal exchange operator, the method captures the dominant nonlocal exchange physics governing band gaps, band-edge states, and defect-related properties of two-dimensional materials at a computational cost comparable to conventional hybrid functionals such as PBE0. Equation~\eqref{eqn:epsmodel} defines the anisotropic screened-exchange functional introduced in this work.

An important advantage of the present approach is that it is formulated fully
within the DFT framework and implemented self-consistently, with the
screened-exchange potential updated during every self-consistent field
iteration together with the Kohn--Sham orbitals until convergence.
As a result, total energies, atomic forces, and structural relaxations are directly accessible at the same computational cost as in hybrid-functional calculations. This capability is essential for the investigation of defects, structural distortions, and geometry-dependent electronic properties in 2D materials. In contrast, while GW provides highly accurate quasiparticle energies, it is not routinely available for total-energy-based structural optimization or force calculations in practical large-scale applications, and its computational cost is substantially higher.
The present functional is intended for thin two-dimensional materials whose dielectric screening can be described by the effective quasi-two-dimensional interaction employed here. It is therefore applicable to free-standing as well as environmentally screened systems.

\section{Results}

In this section, we assess the performance of the anisotropic screened-exchange
functional introduced in Sec.~\ref{sec:theory} for representative
2D semiconductors.
The validation focuses on three key aspects that are directly governed by dielectric screening in reduced dimensionality: \\
(i) quasi-particle band gaps, \\
(ii) the dispersion of band-edge states relevant for defect physics, and \\
(iii) the piecewise-linear behavior of the total energy with respect to fractional occupation numbers. \\
In addition, we examine the suitability of the resulting electronic structure as a starting point for optical calculations in the linear response TDDFT approach.

All calculations were performed using a modified version of the Vienna \emph{ab initio} Simulation Package (\textsc{Vasp}~5.4.4)~\cite{VASP:3,VASP:4} within the projector-augmented-wave framework, treating semicore $d$ states as valence electrons. The implementation of the anisotropic screened-exchange functional follows directly from the formalism described above.
These modifications of the \textsc{Vasp} source code can be made available to certified owners of a \textsc{Vasp} 
user license, once they have been ported to \textsc{Vasp} 6.

\begin{figure}[!ht]
\centering
\includegraphics[width=0.6\textwidth,angle=0]{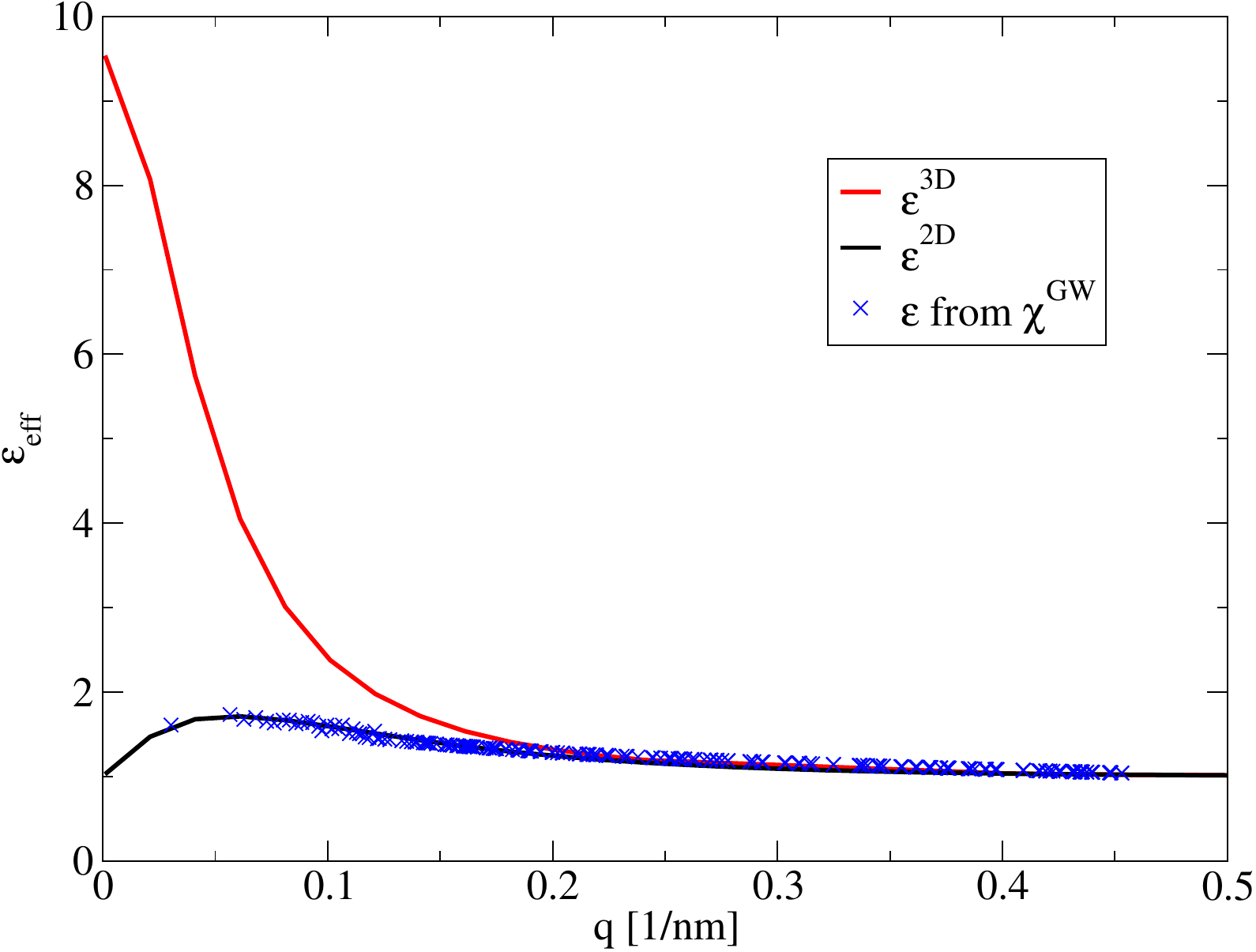}
\caption{
Wave-vector-dependent macroscopic dielectric screening for monolayer GaSe.
The red curve shows the isotropic three-dimensional screening function used in
the bulk screened-exchange functional of Ref.~\cite{Lorke20Koopmans}, where the
screened interaction depends only on the magnitude of the wave vector through
a scalar dielectric function.
The black curve shows the anisotropic two-dimensional screening introduced in
the present work, Eq.~\eqref{eqn:epsmodel}, where the effective interaction
explicitly accounts for the different in-plane and out-of-plane dielectric
response of a finite-thickness monolayer.
Blue crosses denote the static dielectric response extracted independently
from the microscopic response function calculated within the GW$_0$/RPA
framework.
The close agreement between the present anisotropic screening model and the
GW$_0$ response is the expected result and validates the screening entering
the exchange--correlation functional.
In contrast, the deviation from the isotropic bulk screening demonstrates the
importance of incorporating the correct two-dimensional screening behavior.}
\label{fig1}
\end{figure}

The effective macroscopic screening obtained from the anisotropic screened-exchange model is shown in Fig.~\ref{fig1} for monolayer GaSe as a representative example. 
In contrast to bulk materials, where screening approaches a constant macroscopic dielectric constant in the long-wavelength limit, the 2D screening function satisfies $\varepsilon(q_{\parallel}\!\rightarrow\!0)=1$ and also approaches unity for large wave vectors. 
This behavior reflects the absence of macroscopic polarization perpendicular to the layer and the nonlocal character of dielectric screening in 2D systems.

Calculations on the unit cell of the  materials were performed 
using a 12$\times$12$\times$1 $\Gamma$-centered 
Monkhorst-Pack \cite{Monkhorst:76} grid. For defect 
calculations, 128 (144,108) atom supercells were used for hBN (GaSe,MoS$_2$), 
applying the $\Gamma$-point approximation.
The defect geometries were fully relaxed.
A 450eV (900eV) cutoff was applied for the expansion of the wave functions (charge density).
The computational cost for practical calculations of defects and adsorbates is on par with that of DFT calculations
with the PBE0 exchange potential.
Charge corrections for the total energy were performed by the slabcc method \cite{Tabriz:2019}.
As no direct analogue to the method of Chen and Pascarello \cite{Chen:13} for the correction of single-particle energies exists for two-dimensional materials and the respective total energy corrections, we opt to investigate the piecewise linearity directly.
As a reference for the quasi-particle band gaps, we perform GW$_0$ calculations starting from PBE wave functions, using 1000 bands and a 12$\times$12$\times$1 $\Gamma$-centered Monkhorst-Pack grid. These calculations also provide an estimate of the quasi-particle renormalization factor $Z$.
In the present calculations, the experimental lattice constants are used for consistency with the literature.

\begin{figure}[h!]
  \begin{center}
    \includegraphics[width=0.6\textwidth,angle=0]{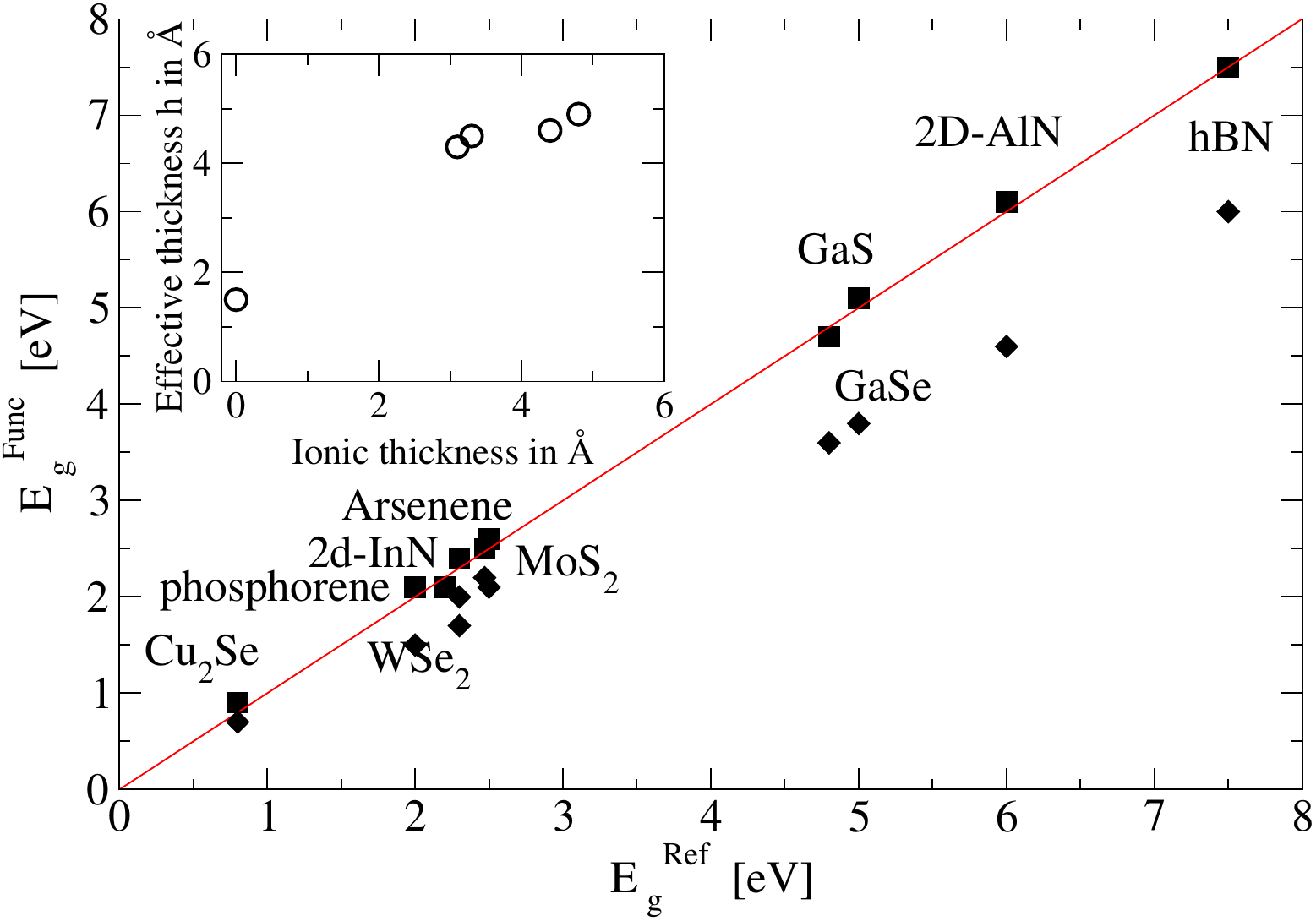}
    \caption{
Quasi-particle band gaps E$_G^\text{Func}$ of representative 2D 
materials as a function of the reference GW$_0$ band gap.
Squares denote results obtained with the present anisotropic screened-exchange 
functional.
Diamonds correspond to results obtained using the isotropically screened 
three-dimensional functional of Ref.~\cite{Lorke20Koopmans}.
The materials shown were selected to span a broad range of band gaps, bonding 
character, and dielectric screening strengths typical of 2D 
semiconductors.
The circles in the inset indicate the effective layer thickness as a function 
of ionic thickness for the material.
The anisotropic screened-exchange functional systematically improves band-gap 
predictions by incorporating the correct 2D screening behavior, 
yielding significantly better agreement with GW$_0$ results compared to the 
isotropic approach.
\label{gap}}
  \end{center}
\end{figure}

In Fig.~\ref{gap} we show the band gaps obtained with the present anisotropic 
screened-exchange functional for representative freestanding 2D 
semiconductors spanning a wide range of electronic structure characteristics, 
from small-gap systems such as Cu$_2$Se to wide-gap insulators such as hBN. 
Overall, the functional reproduces the GW$_0$ reference band gaps with good 
accuracy across this diverse set of materials.

The agreement is particularly good for main-group semiconductors where the 
band edges are dominated by $s$- and $p$-derived states. Slightly larger 
deviations are observed for transition-metal dichalcogenides such as 
MoS$_2$ and WSe$_2$, which is consistent with the known limitations of 
screened-exchange approaches in systems with partially filled $d$ shells. 
These deviations remain moderate and do not affect the qualitative agreement 
with GW$_0$.

Importantly, the functional is not constructed by fitting material-specific 
parameters, but instead derives its screening behavior from a physically 
motivated macroscopic model of 2D dielectric screening. 
Consequently, perfect agreement for every individual compound is not expected. 
Rather, the results demonstrate that incorporating anisotropic screening 
systematically captures the dominant quasi-particle corrections across 
representative classes of 2D semiconductors.

For the particularly relevant cases of hBN, GaSe, and GaS, the inset of Fig.~\ref{gap} shows the effective layer thickness $h$ entering the macroscopic screening model. 
These values are consistent with the physical extent of the electronic density perpendicular to the layer. 
In particular, for hBN the effective electronic thickness slightly exceeds the ionic layer thickness, reflecting the spatial extent of the valence orbitals that contribute to dielectric screening.

To isolate the role of anisotropic screening, we also evaluated band gaps using the isotropic three-dimensional screened-exchange functional of Ref.~\cite{Lorke20Koopmans}. 
While both approaches yield similar results for small-gap materials such as Cu$_2$Se, substantial deviations emerge for strongly 2D systems, including hBN and the transition-metal dichalcogenides. 
This comparison demonstrates that incorporating anisotropic dielectric screening is essential for accurately describing quasi-particle band gaps in 2D materials.

While the present benchmark set focuses on nonmagnetic semiconducting monolayers,
it covers a broad range of band gaps, chemical bonding situations, and dielectric
screening strengths relevant for defect physics in 2D materials.
Like the bulk functional of Ref.~\cite{Lorke20Koopmans}, the present functional is
transferable between compounds of the same class; for example, alloys or Janus
materials such as MoSSe will be tractable within the same framework.
Perfect agreement for every individual material is therefore not expected, as the
functional is not fitted to specific compounds but instead derived from a
physically motivated screening model.
Nevertheless, the results demonstrate that the approach captures the dominant
quasi-particle corrections across representative classes of 2D
semiconductors.
Small deviations for materials such as MoS$_2$ and WSe$_2$ are expected,
particularly in systems with partially filled $d$ shells, where even bulk
screened-exchange approaches are known to exhibit limitations.

Since macroscopic dielectric screening is determined primarily by the charge response, the screening model of Eq.~\eqref{eqn:epsmodel} does not rely on a specific spin configuration and is in principle expected to remain
valid to leading order also in spin-polarized systems. However, magnetic 2D systems involve additional spin-dependent exchange effects and, in some cases, localized $d$-electron correlations that may require an explicit spin-dependent extension of the screened-exchange operator. Hence, the present work focuses on nonmagnetic semiconducting monolayers. Extending the anisotropic screening framework to magnetic systems represents a natural direction for future investigation.



In many cases, not only the band-gap itself is of importance for a proper description of defect states 
but also the band edges over the whole Brillouin zone, as defect states are often superpositions of all band-edge states.

\begin{figure}[h]
\centering
\includegraphics[width=0.6\textwidth,angle=0]{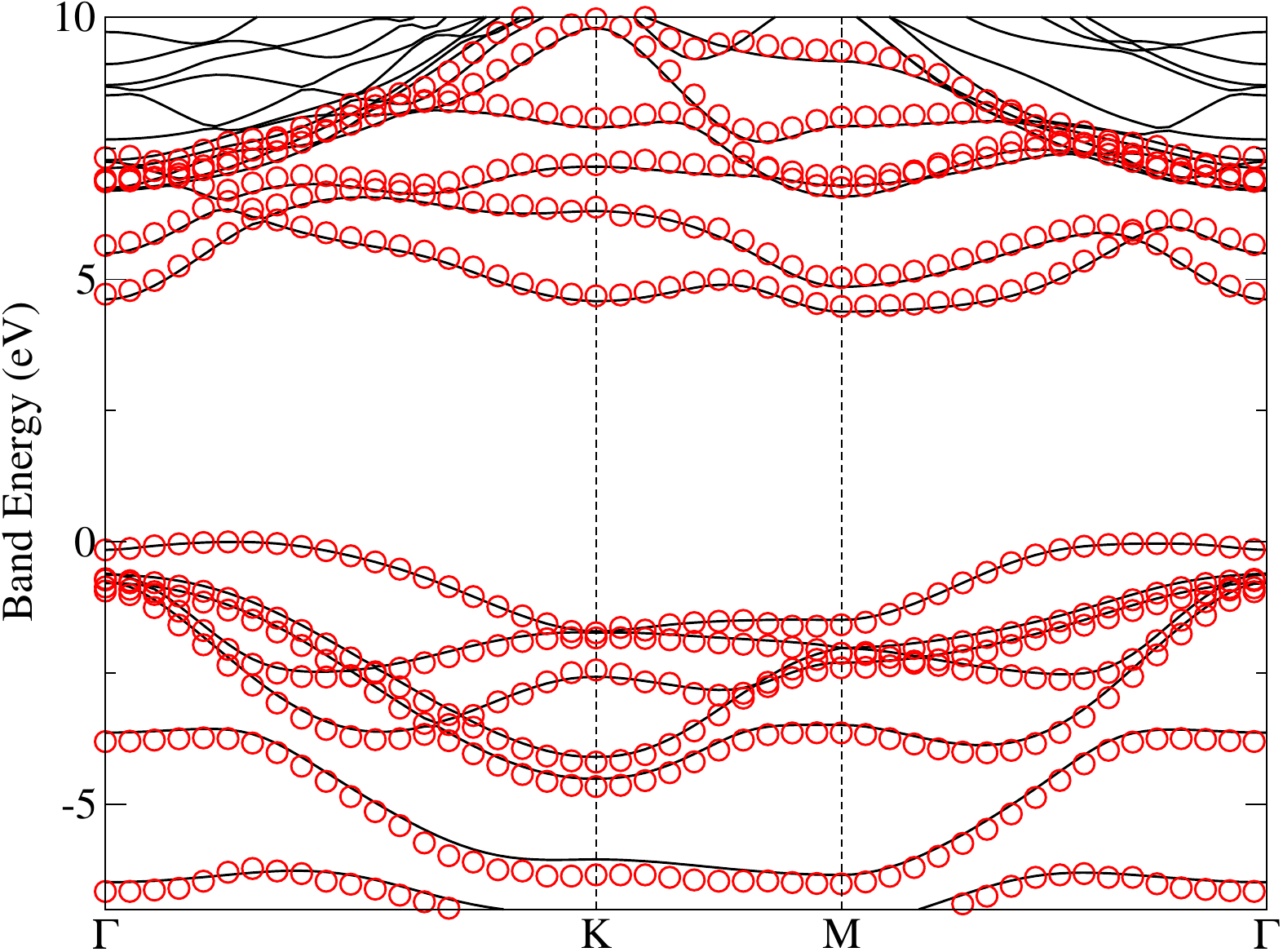}
\caption{
Comparison of the band-edge electronic structure of a representative 2D semiconductor calculated using the present anisotropic screened-exchange functional (circles) and the GW$_0$ approximation (solid lines).
Within this energy window, the anisotropic screened-exchange functional reproduces the GW$_0$ band gap and the dispersion of the band-edge states, demonstrating that the physically motivated anisotropic screening captures the dominant exchange effects relevant for low-energy electronic and defect-related properties. The energy of the valence band maximum has been set to 0.
\label{fig_band}}
\end{figure}

Figure~\ref{fig_band} compares the electronic band structure obtained from the present anisotropic screened-exchange functional with reference GW$_0$ calculations for GaSe. While the primary objective of the functional is an accurate description of the band-edge electronic structure governing defect formation and low-energy excitations, the comparison has been extended over a wider energy range to assess the overall quality of the band dispersion.

In addition to reproducing the quasiparticle gap, the present functional accurately reproduces the overall dispersion of the valence and conduction bands, including their bandwidths and relative band alignment along the high-symmetry path.

In this energy window, containing the first conduction and valence bands, the anisotropic screened-exchange functional yields rather close agreement with GW$_0$ for both the fundamental band gap and the dispersion of the band-edge states.
This indicates that the nonlocal exchange potential, combined with physically motivated anisotropic screening, captures the dominant screening effects relevant for low-energy excitations and defect-related physics.
It is noteworthy that, for some bands, such as the second and third conduction bands, the agreement is better at the high-symmetry points than along the connecting paths.
As expected for a static screened-exchange approximation, the agreement gradually deteriorates for higher conduction states and deeper valence bands, where dynamical self-energy effects become increasingly important. The detailed description of these states is therefore beyond the primary scope of the present functional.
Nevertheless, the level of agreement achieved for multiple bands across the Brillouin zone supports the use of the proposed functional as an efficient and physically motivated approximation to GW for 2D materials.

As expected for a static screened-exchange approximation, quantitative deviations may occur for higher-lying conduction states and deeper valence bands, where dynamical self-energy effects become increasingly important. Nevertheless, the excellent agreement obtained over a broad energy window indicates that the proposed functional provides an efficient and physically well-founded approximation to GW$_0$ for electronic-structure calculations in two-dimensional materials.

\begin{figure}[h]
\centering
\includegraphics[width=0.6\textwidth,angle=0]{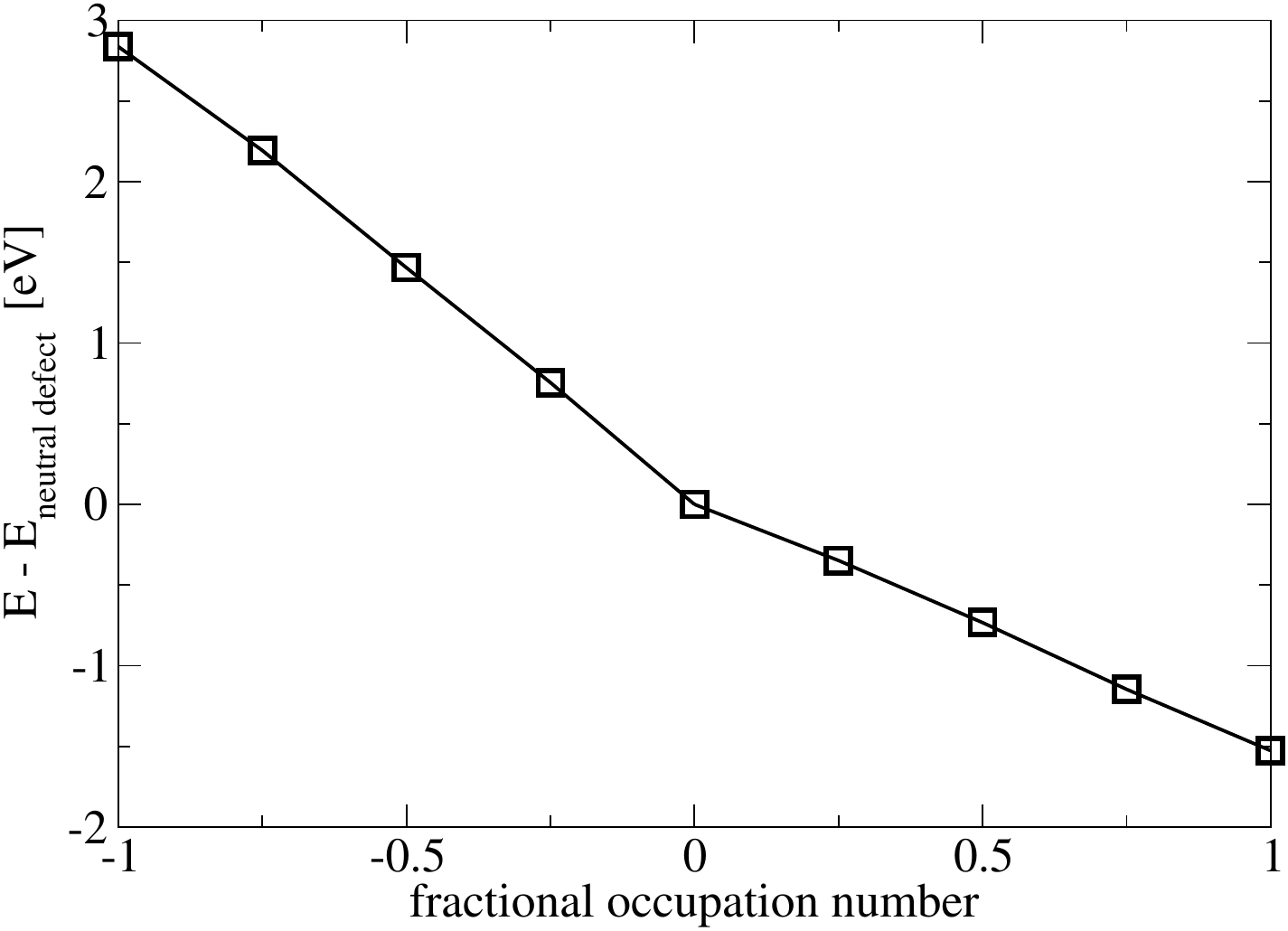}
\caption{Total energy of the $\text{Ge}_\text{Ga}$ substitutional defect in GaSe as a function of the fractional occupation number.
The energy of the neutral defect has been set to 0.
\label{linearity}}
\end{figure}

The piecewise linearity of the total energy with respect to fractional occupation numbers is shown in Fig.~\ref{linearity} for representative substitutional defects in 2D materials.
As an example, Fig.~\ref{linearity} shows the total energy of the $\text{Ge}_\text{Ga}$ substitutional defect in GaSe as a function of fractional occupation of the defect level.

A quadratic fit of the form $E(x)=b_0+b_1 x+b_2 x^2$ yields a curvature of $b_2=0.03$\,eV, indicating near piecewise-linear behavior.
Comparable values of $b_2=0.04$\,eV and $0.08$\,eV are obtained for the $\mathrm{C}_{\mathrm{N}}$ substitutional defect in hBN and the $\mathrm{Nb}_{\mathrm{Mo}}$ substitutional defect in MoS$_2$, respectively.
These small deviations from linearity demonstrate that the present anisotropic screened-exchange functional largely suppresses self-interaction errors and artificial charge delocalization.

The near-linear behavior of the total energy indicates that the anisotropic screened-exchange functional largely restores the generalized Koopmans condition for localized defect states in 2D materials. 
By accurately reproducing both the band-edge electronic structure and near piecewise-linear total-energy behavior, the present approach provides a reliable and computationally efficient framework for describing localized defect states in 2D materials.

\begin{figure}[!h]
\centering
\includegraphics[width=0.6\textwidth,angle=0]{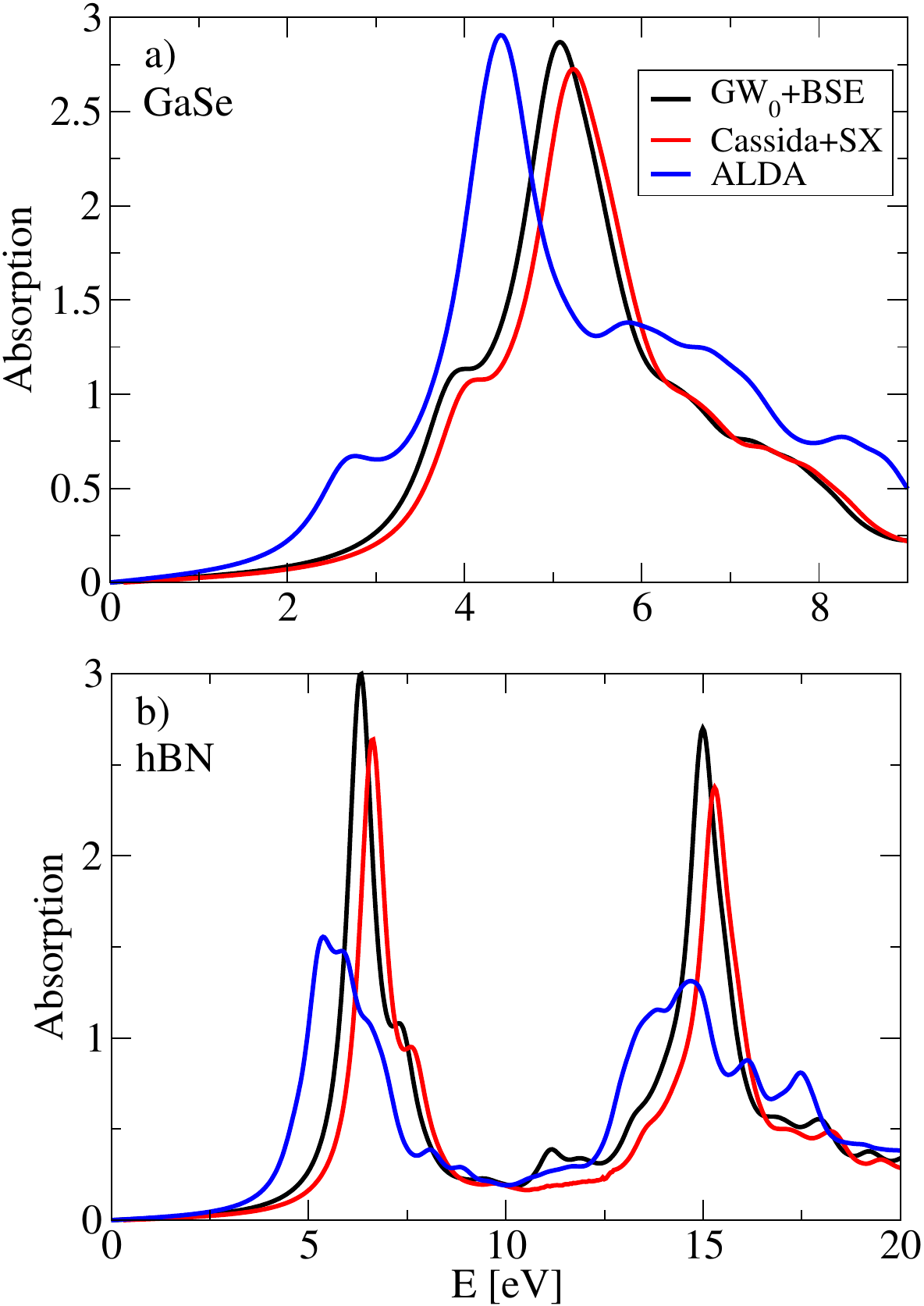}
\caption{Optical absorption from linear response TDDFT, i.e. the Cassida equation, with the functional presented here as Kernel (red) and ALDA (blue). For reference, GW$_0$+BSE (black) calculations are shown. Presented are the spectra for (a) GaSe and (b) hBN.
\label{optics}}
\end{figure}

The optical absorption spectra shown in Fig.~\ref{optics} for GaSe and hBN are computed using two complementary approaches: the GW$_0$ plus Bethe--Salpeter equation (GW$_0$+BSE) framework and linear-response time-dependent density functional theory (Cassida equation).
The GW$_0$+BSE calculations are performed following established best-practice procedures, using PBE as the starting point for the partially self-consistent GW$_0$ calculation.
In the TDDFT calculations, the anisotropic screened-exchange functional is employed consistently both for the ground-state electronic structure and for the exchange--correlation kernel entering the linear-response equations (Casida formalism).
As a result, the long-range electron--hole interaction responsible for excitonic effects in 2D materials is treated in a manner consistent with the underlying anisotropic dielectric screening.

The comparison in Fig.~\ref{optics} shows that the TDDFT spectra obtained with the present functional reproduce the main features of the GW$_0$+BSE results for both GaSe and hBN, including exciton binding energies and relative oscillator strengths.
Some fine spectral features, such as higher-energy shoulders, are not captured with the same accuracy, which is expected within a static screened-exchange framework. 
In contrast, widely used local and adiabatic approximations to the exchange-correlation kernel, most notably the adiabatic local density approximation (ALDA) \cite{RungeGross1984,Casida1995}, are inherently limited by their lack of spatial non-locality and frequency dependence. As a consequence, they fail to capture the long-range electron-hole interaction that is essential for an accurate description of excitonic effects, leading to a systematic underestimation of 
exciton binding energies and significant errors in the resulting optical transition energies.
We emphasize that the spectra shown here are not fully converged with respect to Brillouin-zone sampling; instead, identical $k$-point grids are used for both GW$_0$+BSE and TDDFT calculations to ensure a consistent and meaningful comparison.
These results demonstrate that the anisotropic screened-exchange functional provides a consistent starting point for optical calculations in 2D materials, using the same functional from the electronic structure calculation to the optical properties.



\section{Conclusion}
In conclusion, we have introduced a self-consistent  
anisotropic screened-exchange functional
for two-dimensional 
materials that incorporates wave-vector-dependent dielectric screening 
directly into the Kohn--Sham framework.
The functional derives its screening behavior from a physically motivated 
macroscopic model of reduced-dimensional dielectric response and does not rely 
on material-specific fitting parameters.

Across a representative set of 2D semiconductors, the approach 
reproduces quasi-particle band gaps obtained from GW$_0$ calculations with good 
accuracy, captures the dispersion of band-edge states relevant for defect 
physics, and exhibits near piecewise-linear total-energy behavior with respect 
to fractional occupation numbers. 
The latter indicates a substantial suppression of self-interaction errors and 
a restoration of generalized Koopmans behavior for localized defect states. 
In addition, the functional provides a consistent and computationally efficient 
starting point for optical calculations within linear-response TDDFT, yielding 
optical spectra in good agreement with GW$_0$+BSE results for prototypical 
2D materials.
Because the present functional is formulated within the Kohn--Sham framework, it provides direct access to total energies, forces, and structural relaxations while incorporating physically motivated anisotropic screened exchange. This enables predictive simulations of defects, structural properties, and optical excitations in 2D materials at a computational cost comparable to hybrid functionals. The present work therefore establishes an efficient and physically grounded framework for the first-principles investigation and design of 2D materials.

At present, the functional is primarily designed for nonmagnetic semiconducting monolayers. 
Although the underlying dielectric screening model is formulated at the level of 
charge response, magnetic systems may introduce additional spin-dependent 
exchange effects that need further investigation. 
Extending the anisotropic screening framework to spin-polarized 2D
materials represents a natural direction for future work.

Overall, the results demonstrate that the incorporation of the correct 
anisotropic dielectric screening into a static screened-exchange framework 
provides an efficient and physically transparent approximation to quasi-particle 
methods for 2D materials, particularly in the context of defect 
and optoelectronic properties.

\section{Acknowledgment}
M.L acknowledges funding by the Deutsche Forschungsgemeinschaft (DFG, German Research Foundation) under grant number LO 1840/7-1 as well as support from the DFG via SFB 1242 (Project-ID 278162697) and CPU time at HLRN (Berlin/G\"ottingen).
Stimulating discussions with A. Steinhoff (U Bremen), P. Kratzer (U Duisburg-Essen) and E.K.U. Gross (U Jerusalem) are gratefully acknowledged.
\bibliography{refs_ml,references_paper_neu}

@Article{VASP:3,
  Title                    = {Efficiency of ab-initio total energy calculations for metals and semiconductors using a plane-wave basis set},
  Author                   = {Kresse, G. and Furthm\"uller, J.},
  Journal                  = {Comput. Mat. Sci.},
  Year                     = {1996},

  Month                    = {July},
  Number                   = {1},
  Pages                    = {15-50},
  Volume                   = {6}
}

@Article{VASP:4,
  Title                    = {Efficient iterative schemes for ab initio total-energy calculations using a plane-wave basis set},
  Author                   = {Kresse, G. and Furthm\"uller, J.},
  Journal                  = {Phys. Rev. B},
  Year                     = {1996},
  Pages                    = {11169},
  Volume                   = {54}
}

@Article{HSE,
  author = 	 {J. Heyd and G. E. Scuseria  and M. Ernzerhof},
  title = 	 {},
  journal = 	 {J. Chem. Phys.},
  year = 	 {2003},
  OPTkey = 	 {},
  volume = 	 {118},
  OPTnumber = 	 {},
  pages = 	 {8207},
  OPTmonth = 	 {},
  OPTnote = 	 {},
  OPTannote = 	 {}
}

@Article{PBE0,
  author = 	 {C. Adamo and V. Barone},
  title = 	 {},
  journal = 	 {J. Chem. Phys.},
  year = 	 {1999},
  OPTkey = 	 {},
  volume = 	 {110},
  OPTnumber = 	 {},
  pages = 	 {6158},
  OPTmonth = 	 {},
  OPTnote = 	 {},
  OPTannote = 	 {}
}

@article{Shimazaki:08,
title = "Band structure calculations based on screened Fock exchange method",
journal = "Chemical Physics Letters",
volume = "466",
number = "1",
pages = "91 - 94",
year = "2008",
issn = "0009-2614",
doi = "https://doi.org/10.1016/j.cplett.2008.10.012",
url = "http://www.sciencedirect.com/science/article/pii/S0009261408013717",
author = "Tomomi Shimazaki and Yoshihiro Asai",
}

@article{Cappellini:93,
  title = {Model dielectric function for semiconductors},
  author = {Cappellini, G. and Del Sole, R. and Reining, Lucia and Bechstedt, F.},
  journal = {Phys. Rev. B},
  volume = {47},
  issue = {15},
  pages = {9892--9895},
  numpages = {0},
  year = {1993},
  month = {Apr},
  publisher = {American Physical Society},
  doi = {10.1103/PhysRevB.47.9892},
  url = {https://link.aps.org/doi/10.1103/PhysRevB.47.9892}
}

@article{Shimazaki:10,
author = {Shimazaki,Tomomi  and Asai,Yoshihiro },
title = {Energy band structure calculations based on screened Hartree--Fock exchange method: Si, AlP, AlAs, GaP, and GaAs},
journal = {The Journal of Chemical Physics},
volume = {132},
number = {22},
pages = {224105},
year = {2010},
doi = {10.1063/1.3431293},
URL = { 
        https://doi.org/10.1063/1.3431293
    },
eprint = { 
        https://doi.org/10.1063/1.3431293
    }
}

@article{Chen:18,
  title = {Nonempirical dielectric-dependent hybrid functional with range separation for semiconductors and insulators},
  author = {Chen, Wei and Miceli, Giacomo and Rignanese, Gian-Marco and Pasquarello, Alfredo},
  journal = {Phys. Rev. Materials},
  volume = {2},
  issue = {7},
  pages = {073803},
  numpages = {14},
  year = {2018},
  month = {Jul},
  publisher = {American Physical Society},
}

@article{Deak:17,
  title = {Choosing the correct hybrid for defect calculations: A case study on intrinsic carrier trapping in $\ensuremath{\beta}\ensuremath{-}\mathrm{G}{\mathrm{a}}_{2}{\mathrm{O}}_{3}$},
  author = {De\'ak, Peter and Duy Ho, Quoc and Seemann, Florian and Aradi, B\'alint and Lorke, Michael and Frauenheim, Thomas},
  journal = {Phys. Rev. B},
  volume = {95},
  issue = {7},
  pages = {075208},
  numpages = {11},
  year = {2017},
  month = {Feb},
  publisher = {American Physical Society},
}

@article{Deak:19,
  title = {Carbon in GaN: Calculations with an optimized hybrid functional},
  author = {De\'ak, Peter and Lorke, Michael and Aradi, B\'alint and Frauenheim, Thomas},
  journal = {Phys. Rev. B},
  volume = {99},
  issue = {8},
  pages = {085206},
  numpages = {8},
  year = {2019},
  month = {Feb},
  publisher = {American Physical Society},
  doi = {10.1103/PhysRevB.99.085206},
  url = {https://link.aps.org/doi/10.1103/PhysRevB.99.085206}
}

@article{Janak:78,
  title = {Proof that $\frac{\ensuremath{\partial}E}{\ensuremath{\partial}{n}_{i}}=\ensuremath{\epsilon}$ in density-functional theory},
  author = {Janak, J. F.},
  journal = {Phys. Rev. B},
  volume = {18},
  issue = {12},
  pages = {7165--7168},
  numpages = {0},
  year = {1978},
  month = {Dec},
  publisher = {American Physical Society},
  doi = {10.1103/PhysRevB.18.7165},
  url = {https://link.aps.org/doi/10.1103/PhysRevB.18.7165}
}

@article{Chen:13,
  title = {Correspondence of defect energy levels in hybrid density functional theory and many-body perturbation theory},
  author = {Chen, Wei and Pasquarello, Alfredo},
  journal = {Phys. Rev. B},
  volume = {88},
  issue = {11},
  pages = {115104},
  numpages = {8},
  year = {2013},
  month = {Sep},
  publisher = {American Physical Society},
  doi = {10.1103/PhysRevB.88.115104},
  url = {https://link.aps.org/doi/10.1103/PhysRevB.88.115104}
}

@article{Lany:09,
  title = {Polaronic hole localization and multiple hole binding of acceptors in oxide wide-gap semiconductors},
  author = {Lany, Stephan and Zunger, Alex},
  journal = {Phys. Rev. B},
  volume = {80},
  issue = {8},
  pages = {085202},
  numpages = {5},
  year = {2009},
  month = {Aug},
  publisher = {American Physical Society},
  doi = {10.1103/PhysRevB.80.085202},
  url = {https://link.aps.org/doi/10.1103/PhysRevB.80.085202}
}

@article{Clark:10,
  title = {Screened exchange density functional applied to solids},
  author = {Clark, Stewart J. and Robertson, John},
  journal = {Phys. Rev. B},
  volume = {82},
  issue = {8},
  pages = {085208},
  numpages = {11},
  year = {2010},
  month = {Aug},
  publisher = {American Physical Society},
  doi = {10.1103/PhysRevB.82.085208},
  url = {https://link.aps.org/doi/10.1103/PhysRevB.82.085208}
}

@article{Clark:10a,
  title = {Intrinsic defects in ZnO calculated by screened exchange and hybrid density functionals},
  author = {Clark, S. J. and Robertson, J. and Lany, S. and Zunger, A.},
  journal = {Phys. Rev. B},
  volume = {81},
  issue = {11},
  pages = {115311},
  numpages = {5},
  year = {2010},
  month = {Mar},
  publisher = {American Physical Society},
  doi = {10.1103/PhysRevB.81.115311},
  url = {https://link.aps.org/doi/10.1103/PhysRevB.81.115311}
}

@article{Zheng:11,
  title = {Improving Band Gap Prediction in Density Functional Theory from Molecules to Solids},
  author = {Zheng, Xiao and Cohen, Aron J. and Mori-S\'anchez, Paula and Hu, Xiangqian and Yang, Weitao},
  journal = {Phys. Rev. Lett.},
  volume = {107},
  issue = {2},
  pages = {026403},
  numpages = {4},
  year = {2011},
  month = {Jul},
  publisher = {American Physical Society},
  doi = {10.1103/PhysRevLett.107.026403},
  url = {https://link.aps.org/doi/10.1103/PhysRevLett.107.026403}
}

@article{Becke:96,
author = {Becke,Axel D. },
title = {Density functional thermochemistry. IV. A new dynamical correlation functional and implications for exact-exchange mixing},
journal = {The Journal of Chemical Physics},
volume = {104},
number = {3},
pages = {1040-1046},
year = {1996},
doi = {10.1063/1.470829},
}

@article{Monkhorst:76,
  title = {Special points for Brillouin-zone integrations},
  author = {Monkhorst, Hendrik J. and Pack, James D.},
  journal = {Phys. Rev. B},
  volume = {13},
  issue = {12},
  pages = {5188--5192},
  numpages = {0},
  year = {1976},
  month = {Jun},
  publisher = {American Physical Society},
  doi = {10.1103/PhysRevB.13.5188},
  url = {https://link.aps.org/doi/10.1103/PhysRevB.13.5188}
}

@article{Perdew:81,
  title = {Self-interaction correction to density-functional approximations for many-electron systems},
  author = {Perdew, J. P. and Zunger, Alex},
  journal = {Phys. Rev. B},
  volume = {23},
  issue = {10},
  pages = {5048--5079},
  numpages = {0},
  year = {1981},
  month = {May},
  publisher = {American Physical Society},
  doi = {10.1103/PhysRevB.23.5048},
  url = {https://link.aps.org/doi/10.1103/PhysRevB.23.5048}
}

@article{Baumeier:06,
  title = {Self-interaction-corrected pseudopotentials for silicon carbide},
  author = {Baumeier, Bj\"orn and Kr\"uger, Peter and Pollmann, Johannes},
  journal = {Phys. Rev. B},
  volume = {73},
  issue = {19},
  pages = {195205},
  numpages = {12},
  year = {2006},
  month = {May},
  publisher = {American Physical Society},
  doi = {10.1103/PhysRevB.73.195205},
  url = {https://link.aps.org/doi/10.1103/PhysRevB.73.195205}
}

@article{Krukau:08,
author = {Krukau,Aliaksandr V.  and Scuseria,Gustavo E.  and Perdew,John P.  and Savin,Andreas },
title = {Hybrid functionals with local range separation},
journal = {The Journal of Chemical Physics},
volume = {129},
number = {12},
pages = {124103},
year = {2008},
doi = {10.1063/1.2978377},
}

@article{Deak:10,
  title = {Accurate defect levels obtained from the HSE06 range-separated hybrid functional},
  author = {De\'ak, Peter and Aradi, B\'alint and Frauenheim, Thomas and Janz\'en, Erik and Gali, Adam},
  journal = {Phys. Rev. B},
  volume = {81},
  issue = {15},
  pages = {153203},
  numpages = {4},
  year = {2010},
  month = {Apr},
  publisher = {American Physical Society},
  doi = {10.1103/PhysRevB.81.153203},
  url = {https://link.aps.org/doi/10.1103/PhysRevB.81.153203}
}

@article{Perdew:87,
  title = {Indirect-path methods for atomic and molecular energies, and new Koopmans theorems},
  author = {Levy, Mel and Pathak, Rajeev K. and Perdew, John P. and Wei, Siqing},
  journal = {Phys. Rev. A},
  volume = {36},
  issue = {5},
  pages = {2491--2494},
  numpages = {0},
  year = {1987},
  month = {Sep},
  publisher = {American Physical Society},
  doi = {10.1103/PhysRevA.36.2491},
  url = {https://link.aps.org/doi/10.1103/PhysRevA.36.2491}
}

@article{Perdew:97a,
  title = {Comment on ``Significance of the highest occupied Kohn-Sham eigenvalue''},
  author = {Perdew, John P. and Levy, Mel},
  journal = {Phys. Rev. B},
  volume = {56},
  issue = {24},
  pages = {16021--16028},
  numpages = {0},
  year = {1997},
  month = {Dec},
  publisher = {American Physical Society},
  doi = {10.1103/PhysRevB.56.16021},
  url = {https://link.aps.org/doi/10.1103/PhysRevB.56.16021}
}

@article{Han:17,
  title = {Defect physics in intermediate-band materials: Insights from an optimized hybrid functional},
  author = {Han, Miaomiao and Zeng, Zhi and Frauenheim, Thomas and De\'ak, Peter},
  journal = {Phys. Rev. B},
  volume = {96},
  issue = {16},
  pages = {165204},
  numpages = {9},
  year = {2017},
  month = {Oct},
  publisher = {American Physical Society},
  doi = {10.1103/PhysRevB.96.165204},
  url = {https://link.aps.org/doi/10.1103/PhysRevB.96.165204}
}

@article{Alkauskas:11,
author = {Alkauskas, Audrius and Broqvist, Peter and Pasquarello, Alfredo},
title = {Defect levels through hybrid density functionals: Insights and applications},
journal = {physica status solidi (b)},
volume = {248},
number = {4},
pages = {775-789},
doi = {10.1002/pssb.201046195},
url = {https://onlinelibrary.wiley.com/doi/abs/10.1002/pssb.201046195},
eprint = {https://onlinelibrary.wiley.com/doi/pdf/10.1002/pssb.201046195},
}

@article{Skone:14,
  title = {Self-consistent hybrid functional for condensed systems},
  author = {Skone, Jonathan H. and Govoni, Marco and Galli, Giulia},
  journal = {Phys. Rev. B},
  volume = {89},
  issue = {19},
  pages = {195112},
  numpages = {12},
  year = {2014},
  month = {May},
  publisher = {American Physical Society},
  doi = {10.1103/PhysRevB.89.195112},
  url = {https://link.aps.org/doi/10.1103/PhysRevB.89.195112}
}

@article{Nguyen:18,
  title = {Koopmans-Compliant Spectral Functionals for Extended Systems},
  author = {Nguyen, Ngoc Linh and Colonna, Nicola and Ferretti, Andrea and Marzari, Nicola},
  journal = {Phys. Rev. X},
  volume = {8},
  issue = {2},
  pages = {021051},
  numpages = {12},
  year = {2018},
  month = {May},
  publisher = {American Physical Society},
  doi = {10.1103/PhysRevX.8.021051},
  url = {https://link.aps.org/doi/10.1103/PhysRevX.8.021051}
}

@article{Perdew:81SI,
  title = {Self-interaction correction to density-functional approximations for many-electron systems},
  author = {Perdew, J. P. and Zunger, Alex},
  journal = {Phys. Rev. B},
  volume = {23},
  issue = {10},
  pages = {5048--5079},
  numpages = {0},
  year = {1981},
  month = {May},
  publisher = {American Physical Society},
  doi = {10.1103/PhysRevB.23.5048},
  url = {https://link.aps.org/doi/10.1103/PhysRevB.23.5048}
}

@article{Dabo:10,
  title = {Koopmans' condition for density-functional theory},
  author = {Dabo, Ismaila and Ferretti, Andrea and Poilvert, Nicolas and Li, Yanli and Marzari, Nicola and Cococcioni, Matteo},
  journal = {Phys. Rev. B},
  volume = {82},
  issue = {11},
  pages = {115121},
  numpages = {16},
  year = {2010},
  month = {Sep},
  publisher = {American Physical Society},
  doi = {10.1103/PhysRevB.82.115121},
  url = {https://link.aps.org/doi/10.1103/PhysRevB.82.115121}
}

@Article{ Banyai:98,
	author = "L. B{\'a}nyai and P. Gartner and H. Haug",
	title = "Self-consistent {RPA} retarded polaron {Green} function for quantum kinetics",
	journal = "Eur. Phys. J. B",
	year = "1998",
	volume = "\textbf{1}",
	pages = "209"
}

@Article{ Gartner:02,
	author = "P. Gartner and L. B{\'a}nyai and H. Haug",
	title = "Self-consistent {RPA} for the intermediate-coupling polaron",
	journal = "Phys. Rev. B",
	year = "2002",
	volume = "\textbf{66}",
	pages = "75205"
}

@Article{ Gartner:00,
	author = "P. Gartner and L. B{\'a}nyai and H. Haug",
	title = "Coulomb screening in the two-time {Keldysh-Green}-function formalism",
	journal = "Phys. Rev. B",
	year = "2000",
	volume = "\textbf{62}",
	pages = "7116"
}

@article{Deak19slab,
  title = {Defect calculations with hybrid functionals in layered compounds and in slab models},
  author = {De\'ak, Peter and Khorasani, Elham and Lorke, Michael and Farzalipour-Tabriz, Meisam and Aradi, B\'alint and Frauenheim, Thomas},
  journal = {Phys. Rev. B},
  volume = {100},
  issue = {23},
  pages = {235304},
  numpages = {8},
  year = {2019},
  month = {Dec},
  publisher = {American Physical Society},
  doi = {10.1103/PhysRevB.100.235304},
  url = {https://link.aps.org/doi/10.1103/PhysRevB.100.235304}
}

@article{Lorke20Koopmans,
  title = {Koopmans-compliant screened exchange potential with correct asymptotic behavior for semiconductors},
  author = {Lorke, Michael and De\'ak, Peter and Frauenheim, Thomas},
  journal = {Phys. Rev. B},
  volume = {102},
  issue = {23},
  pages = {235168},
  numpages = {5},
  year = {2020},
  month = {Dec},
  publisher = {American Physical Society},
  doi = {10.1103/PhysRevB.102.235168},
  url = {https://link.aps.org/doi/10.1103/PhysRevB.102.235168}
}

@article{rosner_wannier_2015,
  title = {Wannier function approach to realistic Coulomb interactions in layered materials and heterostructures},
  author = {R\"osner, M. and \ifmmode \mbox{\c{S}}\else \c{S}\fi{}a\ifmmode \mbox{\c{s}}\else \c{s}\fi{}\ifmmode \imath \else \i \fi{}o\ifmmode \breve{g}\else \u{g}\fi{}lu, E. and Friedrich, C. and Bl\"ugel, S. and Wehling, T. O.},
  journal = {Phys. Rev. B},
  volume = {92},
  issue = {8},
  pages = {085102},
  numpages = {10},
  year = {2015},
  month = {Aug},
  publisher = {American Physical Society},
  doi = {10.1103/PhysRevB.92.085102},
  url = {https://link.aps.org/doi/10.1103/PhysRevB.92.085102}
}

@article{Lopez-Morales:20,
author = {Gabriel I. L\'{o}pez-Morales and Aziza Almanakly and Sitakanta Satapathy and Nicholas V. Proscia and Harishankar Jayakumar and Valery N. Khabashesku and Pulickel M. Ajayan and Carlos A. Meriles and Vinod M. Menon},
journal = {Opt. Mater. Express},
number = {4},
pages = {843--849},
publisher = {Optica Publishing Group},
title = {Room-temperature single photon emitters in cubic boron nitride nanocrystals},
volume = {10},
month = {Apr},
year = {2020},
url = {https://opg.optica.org/ome/abstract.cfm?URI=ome-10-4-843},
doi = {10.1364/OME.386791},
}

@article{hoetger2021gate,
  title={Gate-switchable arrays of quantum light emitters in contacted monolayer MoS2 van der Waals heterodevices},
  author={H{\"o}tger, Alexander and Klein, Julian and Barthelmi, Katja and Sigl, Lukas and Sigger, Florian and Ma{\"a}nner, Wolfgang and Gyger, Samuel and Florian, Matthias and Lorke, Michael and Jahnke, Frank and others},
  journal={Nano Letters},
  volume={21},
  number={2},
  pages={1040--1046},
  year={2021},
  publisher={American Chemical Society}
}

@article{mitterreiter2021role,
  title={The role of chalcogen vacancies for atomic defect emission in MoS2},
  author={Mitterreiter, Elmar and Schuler, Bruno and Micevic, Ana and Hernang{\'o}mez-P{\'e}rez, Daniel and Barthelmi, Katja and Cochrane, Katherine A and Kiemle, Jonas and Sigger, Florian and Klein, Julian and Wong, Edward and Lorke, M and others},
  journal={Nature communications},
  volume={12},
  number={1},
  pages={1--8},
  year={2021},
  publisher={Nature Publishing Group}
}

@article{klein2019site,
  title={Site-selectively generated photon emitters in monolayer {MoS2} via local helium ion irradiation},
  author={Klein, J and Lorke, M and Florian, M and Sigger, F and Sigl, L and Rey, S and Wierzbowski, J and Cerne, J and M{\"u}ller, K and Mitterreiter, E and others},
  journal={Nature communications},
  volume={10},
  number={1},
  pages={1--8},
  year={2019},
  publisher={Nature Publishing Group}
}

@article{Tabriz:2019,
title = {SLABCC: Total energy correction code for charged periodic slab models},
journal = {Computer Physics Communications},
volume = {240},
pages = {101-105},
year = {2019},
issn = {0010-4655},
doi = {https://doi.org/10.1016/j.cpc.2019.02.018},
url = {https://www.sciencedirect.com/science/article/pii/S0010465519300700},
author = {Meisam {Farzalipour Tabriz} and Bálint Aradi and Thomas Frauenheim and Peter Deák},
}

@article{Ramasubramaniam:12,
  author = {Ramasubramaniam, Ashwin},
  title = {Large excitonic effects in monolayers of molybdenum and tungsten dichalcogenides},
  journal = {Phys. Rev. B},
  volume = {86},
  pages = {115409},
  year = {2012},
  doi = {10.1103/PhysRevB.86.115409}
}

@article{Qiu:13,
  author = {Qiu, Diana Y. and da Jornada, Felipe H. and Louie, Steven G.},
  title = {Optical spectrum of MoS2: Many-body effects and diversity of exciton states},
  journal = {Phys. Rev. Lett.},
  volume = {111},
  pages = {216805},
  year = {2013},
  doi = {10.1103/PhysRevLett.111.216805}
}

@article{Komsa:13,
  author = {Komsa, Hannu-Pekka and Pasquarello, Alfredo},
  title = {Finite-size supercell correction schemes for charged defect calculations},
  journal = {Phys. Rev. Lett.},
  volume = {110},
  pages = {095505},
  year = {2013},
  doi = {10.1103/PhysRevLett.110.095505}
}

@article{Cudazzo:11,
  author = {Cudazzo, Pierluigi and Tokatly, I. V. and Rubio, Angel},
  title = {Dielectric screening in two-dimensional insulators: Implications for excitonic and impurity states},
  journal = {Phys. Rev. B},
  volume = {84},
  pages = {085406},
  year = {2011},
  doi = {10.1103/PhysRevB.84.085406}
}

@article{Andersen:15,
  author = {Andersen, Kristian and Latini, Simone and Thygesen, Kristian S.},
  title = {Dielectric genome of van der Waals heterostructures},
  journal = {Nano Lett.},
  volume = {15},
  pages = {4616--4621},
  year = {2015},
  doi = {10.1021/acs.nanolett.5b01196}
}

@article{RungeGross1984,
  author = {Runge, Erich and Gross, E. K. U.},
  title = {Density-Functional Theory for Time-Dependent Systems},
  journal = {Physical Review Letters},
  volume = {52},
  number = {12},
  pages = {997--1000},
  year = {1984},
  doi = {10.1103/PhysRevLett.52.997}
}

@article{Casida1995,
  author = {Casida, Mark E.},
  title = {Time-Dependent Density Functional Response Theory for Molecules},
  journal = {Recent Advances in Density Functional Methods},
  volume = {1},
  pages = {155--192},
  year = {1995},
  publisher = {World Scientific},
  doi = {10.1142/9789812830586_0005}
}

@article{Emelyanenko:08,
  author = {Emelyanenko, Alexandre M. and Boinovich, Ludmila B.},
  title = {The role of surface forces in the wetting of nanostructured surfaces},
  journal = {Journal of Physics: Condensed Matter},
  volume = {20},
  pages = {494227},
  year = {2008},
  doi = {10.1088/0953-8984/20/49/494227}
}

@article{MoriSanchez:08,
  author  = {Mori-S\'anchez, P. and Cohen, A. J. and Yang, W.},
  title   = {Localization and Delocalization Errors in Density Functional Theory and Implications for Band-Gap Prediction},
  journal = {Physical Review Letters},
  volume  = {100},
  pages   = {146401},
  year    = {2008},
  doi     = {10.1103/PhysRevLett.100.146401}
}

@article{Cohen:12,
  author  = {Cohen, A. J. and Mori-S\'anchez, P. and Yang, W.},
  title   = {Challenges for Density Functional Theory},
  journal = {Chemical Reviews},
  volume  = {112},
  number  = {1},
  pages   = {289--320},
  year    = {2012},
  doi     = {10.1021/cr200107z}
}

\end{document}